\documentclass[aps,prl,twocolumn,superscriptaddress,showpacs]{revtex4}
\usepackage{graphicx}
\usepackage{dcolumn}   
\usepackage{bm}        
\usepackage{amssymb}   
\usepackage{dsfont}

\renewcommand{\i}{\mathrm{i}}

\newcommand{\set}[1]{\left \{ #1\right\}}
\newcommand{\abs}[1]{\left \vert #1\right\vert}
\def \tr {\mathrm{Tr}}

\def\A {\mathcal{A}}
\def\X {\mathcal{X}}
\def\M {\mathcal{M}}
\def\mod {\mathrm{mod}}

\usepackage{colordvi}
\usepackage{color}

\begin{document}



\title{Universality in spectral statistics of ``open'' quantum graphs.}
\author{B.~Gutkin} \affiliation{University of Duisburg-Essen, Lotharstr. 1, 47048 Duisburg, Germany}
\author{V.Al.~Osipov} \affiliation{Cologne University, Z\"ulpicher str. 77, 50937 Cologne, Germany}
\vskip 0.25cm

\date{\today}

\begin{abstract}
The  unitary evolution maps in closed chaotic quantum graphs are known to have universal spectral correlations, as predicted by random matrix theory. In chaotic graphs with  absorption  the  quantum  maps  become non-unitary. We show  that their spectral statistics  exhibit  universality at the ``soft'' edges of the spectrum. The same spectral behavior is observed  in many classical  non-unitary  ensembles of random matrices with rotationally invariant measures.

\end{abstract}

\maketitle

  Since 1970's a significant attention of physics community has been  attracted to particularities of the energy spectrum of quantum systems with  chaotic behavior in the classical limit.  In 1984 Bohigas, Giannoni and Schmit (BGS)  conjectured~\cite{BGS1984} that the spectral fluctuations  of closed chaotic Hamiltonian systems are universal and coincide with those of one of three canonical random matrix ensembles (RME).  Based on the semiclassical considerations the validity of BGS conjecture has been established  by now on the physical level of rigor~\cite{Sie, Mul1}.  On the other hand, many  theoretical and experimental studies have focused on open chaotic systems whose  wave dynamics are described by non-unitary evolution operators. Such an opening may occur due to various physical phenomena: attaching external leads  to quantum dots, the dissipation  through the ohmic losses, partial reflection of microwaves at the boundaries of dielectric microcavities etc. It is  of great interest to know whether (and under which conditions) open chaotic systems exhibit universal properties.    So far, the majority  of studies in this respect have been  restricted to  regimes of ``weak'' opening, where the mean dwell time of the particle in the system growth in the semiclassical limit.
For instance, transport properties  of quantum dots with a  finite number of open channels  have been shown to be universal and agree with  the random matrix theory predictions~\cite{Mul2}.
  The main goal of the present paper is  establishing   a new form of spectral universality for   systems with ``strong'' opening when the dwell time remains finite in the semiclassical limit.   

In this work we focus on the model of quantum graphs with broken time reversal symmetry. Quantum graphs were proposed  as a paradigm for the study of compact~\cite{KS1997}  and scattering~\cite{KS2000} quantum chaotic systems. They were also studied 
experimentally  in the presence of absorption~\cite{LHBS2011}. 
Let us  briefly  describe a standard construction of   quantum graphs with $V$ vertices connected by $B$ bonds, see e.g.,~\cite{GS2006} for details. At   bonds $b=1,\dots,B$  the waves $\psi_b$ satisfy  the free Schr\"odinger equation
$\left(-\i{d}_{x_b}+\A_b\right)^2\psi_b(x_b)=k^2\psi_b(x_b)$,
where $x_b\in [0,L_b]$ measures the distance along the bond $b$ and $\A_b$ is a constant vector potential introduced to break the time-reversal symmetry. The corresponding general solution  is a superposition of two plane waves propagating in opposite directions, $\psi_b(x_b)=e^{-\i \A_b x_b} (e^{\i k x_b} a_{b+}+e^{-\i k x_b} a_{b-})$. The constants $a_{b\pm}$  for different bonds  are then connected  in the vertices by means of the scattering matrices $\sigma_v$, $v=1,\dots,V$.
To proceed further one introduces the associated directed graph $\Gamma$  with the double number $N=2B$ of bonds $(b,\bar{b}) $ carrying waves with the positive and negative momenta separately, such that $L_b=L_{\bar{b}}, \A_b=-\A_{\bar{b}}$.   The complete spectral information is carried by the  $N\times N$ \textit{quantum map} $U(k)=S\Lambda(k)$, where  the \textit{scattering   matrix} $S$ depends on the graph's structure and $\sigma_v$'s, $\A_b$'s, while dependence on the energy  $k^2$  is entirely stored in the diagonal  part  $\Lambda(k)=\mathrm{diag}\{e^{ikL_j} \}$, $j\in \{b,\bar{b}\}$. Note that $S$ also fixes  the matrix of ``classical evolution'' $F$ on $\Gamma$, whose elements $F_{i j}= \abs{S_{i\; j}}^2$ specify classical transition probabilities between bonds of the graph.
 The spectrum $\{k_n\}$ of the system is provided by solutions of the secular equation $\det(I-U(k_n))=0$. If all $\sigma_v$ are  unitary and $\A_b$ are real then the resulting  quantum map $U$ is unitary and the system's spectrum is real. It is possible to open  the system by either attaching external leads to the graph or  by introducing absorption at bonds (or vertices) violating the aforementioned conditions. In any such  case  the resulting  (internal) scattering matrix $S$, as well as $U$,  are not unitary anymore and  we will colloquially refer to  $\Gamma$ as \textit{open quantum graph}. 

An appealing feature of quantum graphs is the exact trace formula connecting the density of states $d(k)=\sum_n \delta(k-k_n)$ with the traces of the quantum map $U(k)$.  As a result, the two-point  spectral correlation function can be expressed as  the discrete Fourier transform of the spectral form factor $\langle|\tr U(k)|^2\rangle_k$, where $\langle\cdot\rangle_k\equiv \lim_{K\to\infty}\frac{1}{K}\int_0^K dk(\cdot)$ is the average over the wave number. Furthermore, as was  shown in~\cite{BG2000}, for graphs with  rationally independent bond lengths (which we assume through the paper) the average over $k$ can be traded for the  averages over independent parameters $kL_b$, $b=1,\dots B$. Therefore,  the spectral correlations in individual quantum graphs can be found by solving the same problem  for the ensemble of matrices $U_{\bm \phi}\equiv S\Lambda_{\bm \phi}$,  $\Lambda_{\bm \phi}=\mathrm{diag}\set{e^{\i\phi_k}}_{b=1}^N$, $\phi_b=-\phi_{\bar{b}}$, where averages  $\langle\cdot\rangle_{\bm \phi}$ are taken  over the flat probability measure $\nu(\bm \phi)=\prod_{k=1}^{N/2}\frac{d\phi_k}{2\pi}$. 

The 
 case of graphs with unitary $S$  has been analyzed by both semiclassical \cite{BSW2002, Be2006} and supersymmetry methods \cite{GA2004}. It was demonstrated that under certain condition on the gap of the classical evolution spectrum, BGS conjecture holds i.e., depending on the presence or absence of time reversal symmetry   the ensemble of $U_{\bm \phi}$ has the same spectral statistics as  either Gaussian Unitary or  Gaussian Orthogonal matrices. 
For ``strongly'' open quantum graphs  the   eigenvalues $\{z_k\}^N_{k=1}$ of $U_{\bm \phi}$ are not confined to the unit circle, but rather distributed isotropically  over the complex plane with the mean distance  of the order $1/\sqrt{N}$. (The isotropy follows immediately from the invariance of $\nu(\bm \phi)$ under the rotation
$ \phi_k\to \phi_k+ \phi$ $k=1,\dots N$.)
 Typically, with the increase of  graph's dimension  $z_k$'s become more and more concentrated in an annulus whose boundaries are referred to as \textit{inner} (resp. \textit{outer}) \textit{spectral edge}.   As we show below, a  spectral universality holds at the $1/\sqrt{N}$ neighborhood of these edges. For the sake of simplicity of exposition we  formulate the result for the outer edge and then discuss its extension  to the inner edge.

\textit{Main result.}  
{\sl Let $\Gamma^{N}$ be an infinite  sequence  of open quantum graphs with      $S^{(N)}\Lambda_{\bm \phi}$,  $F^{(N)}$ being   their $N\times N$ quantum  and classical evolution, respectively. For the matrix $F^{(N)}$ we denote by $\lambda$, $\bar{\chi}$, $\chi$  the largest  eigenvalue and the corresponding left (resp. right) eigenvectors normalized by $(\bar{\chi},\chi)=N$.  We will consider the spectrum $\{z_k\}^N_{k=1}$ of rescaled quantum propagator $S\Lambda_{\bm \phi}$, $S\equiv \frac{1}{\sqrt{\lambda}}S^{(N)}$  in the limit $N\to \infty$ under the conditions:
 
(i) Large spectral gap of $F\equiv F^{(N)}/\lambda$: The next to the largest eigenvalue $\lambda_2$ satisfies 
$1-{|\lambda_2|}=\mathcal{O}(N^{-\kappa})$, $\kappa<\frac{1}{2}$;

(ii) Strong non-unitarity of $S$: The parameter $\mu^{(N)}=\frac{1}{N}\tr\left(({S} \X{S}^\dag \bar{\X})^2-(\X\bar{\X})^2\right)$ has a  strictly positive limit $\mu=\lim_{N\to \infty}\mu^{(N)}>0$, where $\X$ ($\bar{\X}$) is the diagonal matrix constructed on $\chi$ ($\bar{\chi}$);

Assuming (i, ii) hold, the spectral density $\tilde{\rho}(z)=\frac{1}{N}\langle\sum_{k=1}^{N}\delta(z-z_k)\rangle_{\bm \phi}$ is a function of $r=\abs{z}$ only
and $\rho(r)\equiv 2\pi r\tilde{\rho}(r) $ has the universal form at $1/\sqrt{N}$ vicinity of the edge $\abs{z}=1$:
\begin{eqnarray}\label{Density}
	\rho\left(1-\frac{s}{\sqrt{N}}\right)=\frac{1}{\mu}\left(2-\mathrm{erfc}\left(\frac{s}{\sqrt{2\mu}}\right)\right)+\mathcal{O}(N^{-\frac{1}{2}+\kappa}),
\end{eqnarray}
in particular $\rho(1)=\mu^{-1}+{o}(N^{0})$. The form-factor $\mathcal{K}(n)=\frac{1}{N}\langle\abs{\tr ( S\Lambda)^n}^2\rangle_{\bm \phi}$ demonstrates the universal asymptotics:
\begin{equation}\label{FormFactor}
\sqrt{N}\mathcal{K}(n)=\frac{2}{ \mu t}\sinh\left( \frac{\mu t^2}{2}\right)+\mathcal{O}(N^{-\frac{1}{2}+\kappa}),
\end{equation}
in the limit where $t=n/\sqrt{N}$ is fixed and $N\to \infty$. 
} 

\begin{figure}
\includegraphics[height=3.1cm]{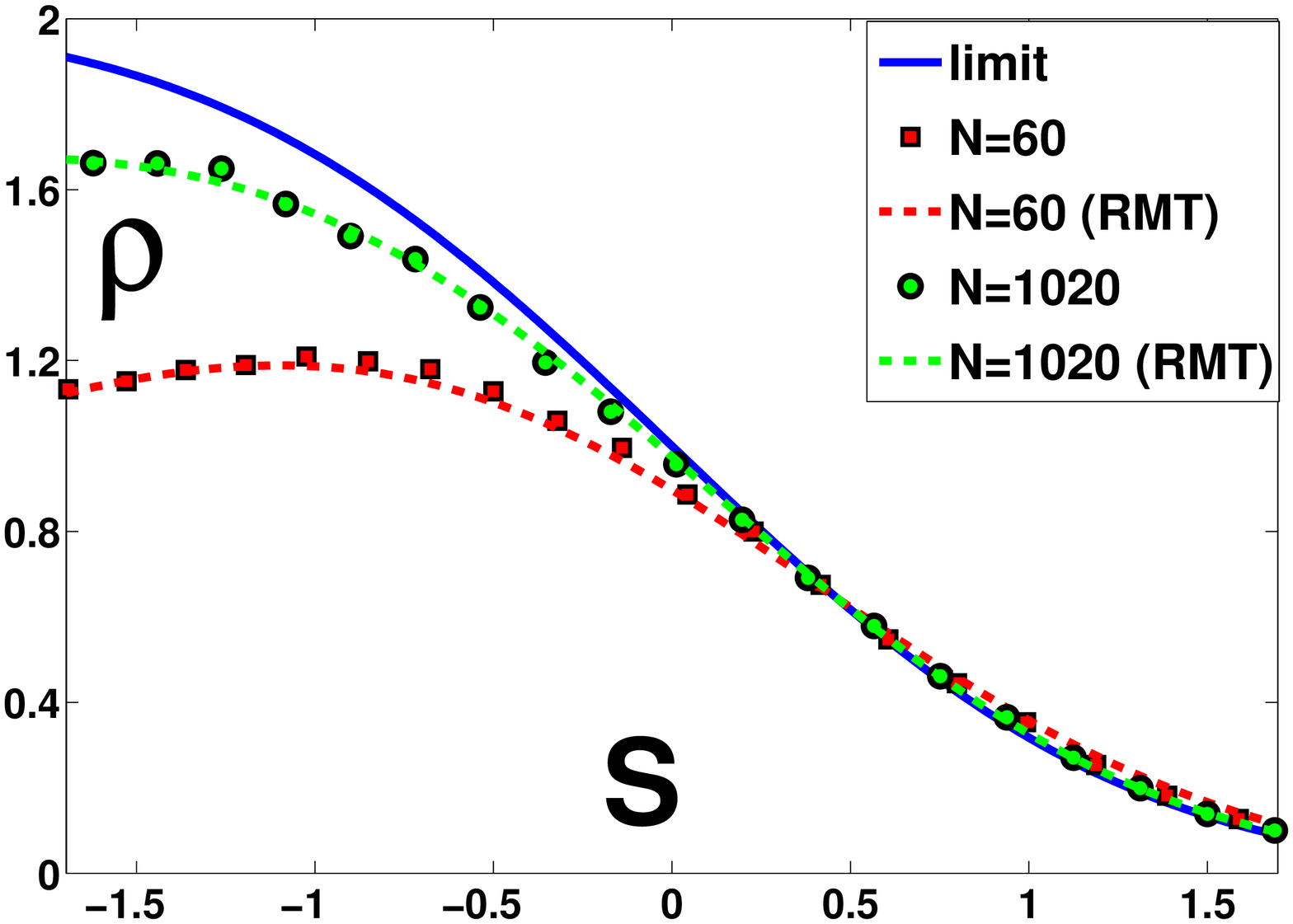}~\hskip-0.5cm~\includegraphics[height=3.1cm]{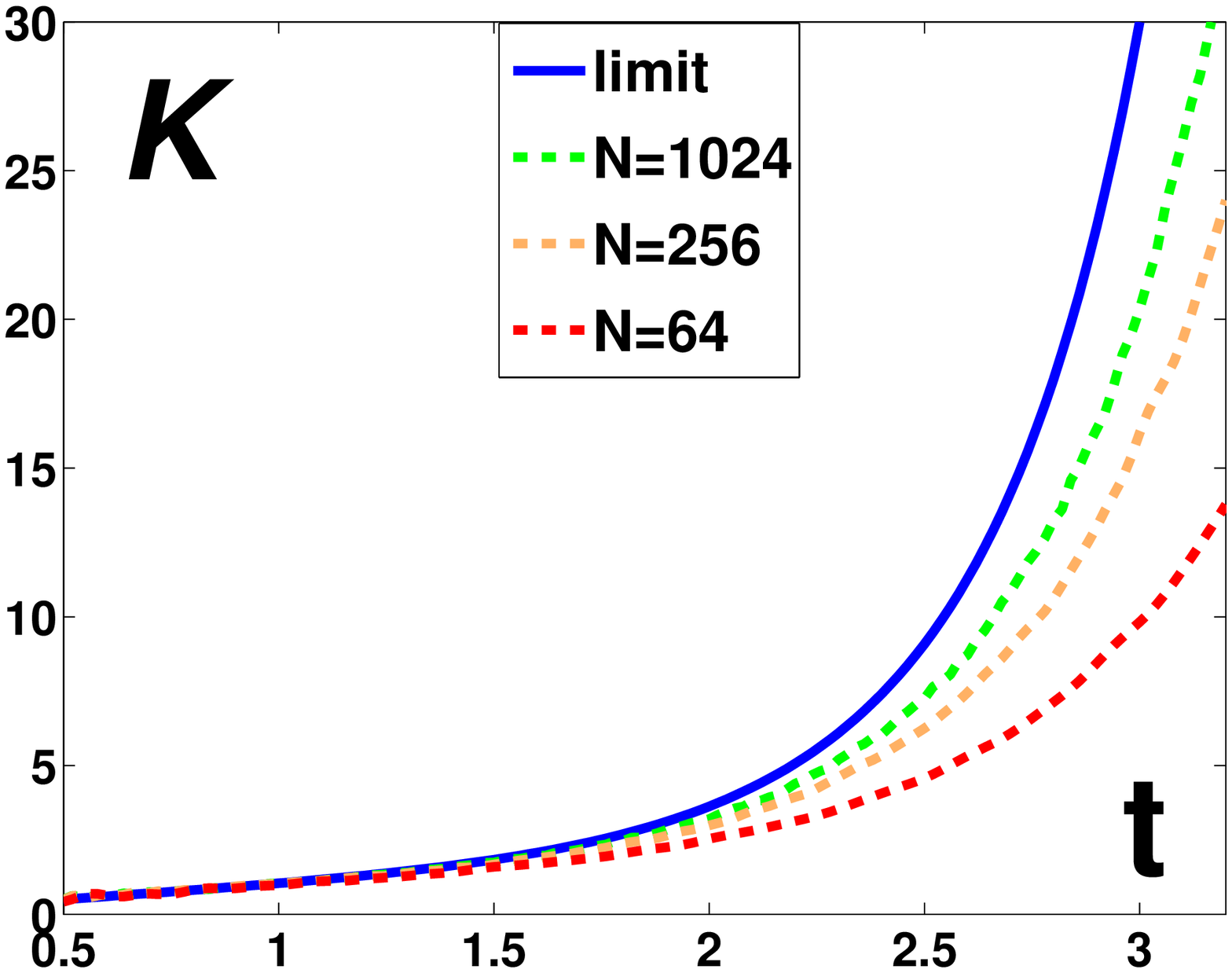} 
\caption{\label{figwide}\small   On the left is the spectral density  of 2-regular connectivity graphs ($\mu=1$) vs. the universal function~(\ref{Density}) -- solid (blue) line; two dashed lines are analytic results from~\cite{ZS2000}  for  RME of truncated unitary matrices of the same dimensions. On the right is  the spectral  form factor for the same family of graphs plotted as dashed lines vs. (\ref{FormFactor}) -- solid (red) line.  }
\end{figure}

Few remarks are in order. 1) Note, that the spectral density of the Ginibre unitary ensemble~\cite{G1965}, and of other ``strongly'' non-unitary ensembles~\cite{ZS2000, B2010} with rotationally invariant measures demonstrate the same soft edge universal form (\ref{Density}). Only the scaling parameter $\mu$ depends on the specifics of these ensembles. 2) Because of the difference in the mean level distance between eigenvalues,
the semiclassical limit $n\sim\sqrt{N}$, $N\to\infty$ considered here differs from the  one in the unitary case $n\sim N$, $N\to\infty$.  
 3) The asymptotics for the  inner edge can be established by considering inverse matrices $(S^{(N)})^{-1}\Lambda^*_{\bm \phi}$ whose  spectrum  $\{z^{-1}_k\}^N_{k=1}$ has the density $\rho'(r)=r^{-2}\rho(1/r)$. This inversion maps the inner edge to the outer  and  (\ref{Density}, \ref{FormFactor}) become applicable to $\rho'$ with the parameter $\mu$ being defined  by the matrices $S^{-1}$. 4)  By eq.~(\ref{Density}) the  outer  and the inner edges  of non-rescaled quantum maps $S^{(N)}\Lambda_{\bm \phi}$ are given by  $\sqrt{\lambda}$ and $1/\sqrt{\lambda'}$,  where $\lambda, \lambda'$ are the highest eigenvalues  of the  ``classical'' maps $|S^{(N)}|^2_{i,j}$, $|(S^{(N)})^{-1}|^2_{i,j}$. If $\lambda'=\infty$ (e.g., $S$ is not invertible) then the inner edge does not exist. 5) The condition (i) on the  spectral gap  of the classical map  is  analogous to Tanner's condition \cite{Tanner} in the unitary case. The difference between  $\kappa<1/2$ (non-unitary)  and $\kappa<1$ (unitary) is due to  the  different time scales involved. It holds for many important classes of graphs, see e.g., \cite{GS2006} and examples below. The condition (ii) implies strong non-unitarity of $S$.  If, for instance,  the number of open channels in a scattering graph is fixed then  $\lim_{N \to \infty}\mu^{(N)}=0$  and  (ii) is violated. Note that $\mu^{(N)}\geq 0$ always and $\mu^{(N)}\equiv 0$ if $S$ is unitary.
\begin{figure}
\begin{center}{
\hskip-0.3cm
\includegraphics[height=3.1cm]{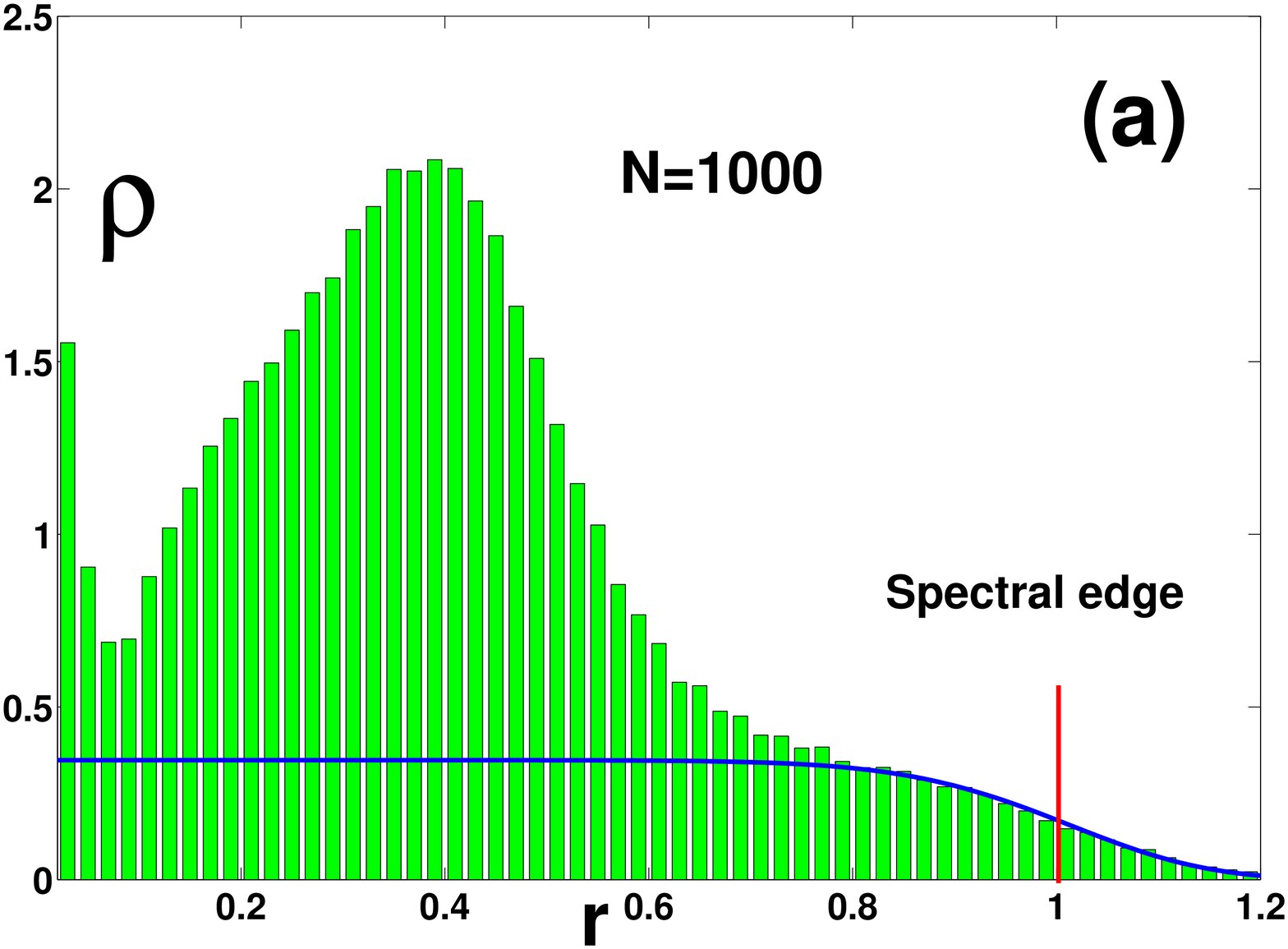}~\hskip-0.5cm~\includegraphics[height=3.1cm]{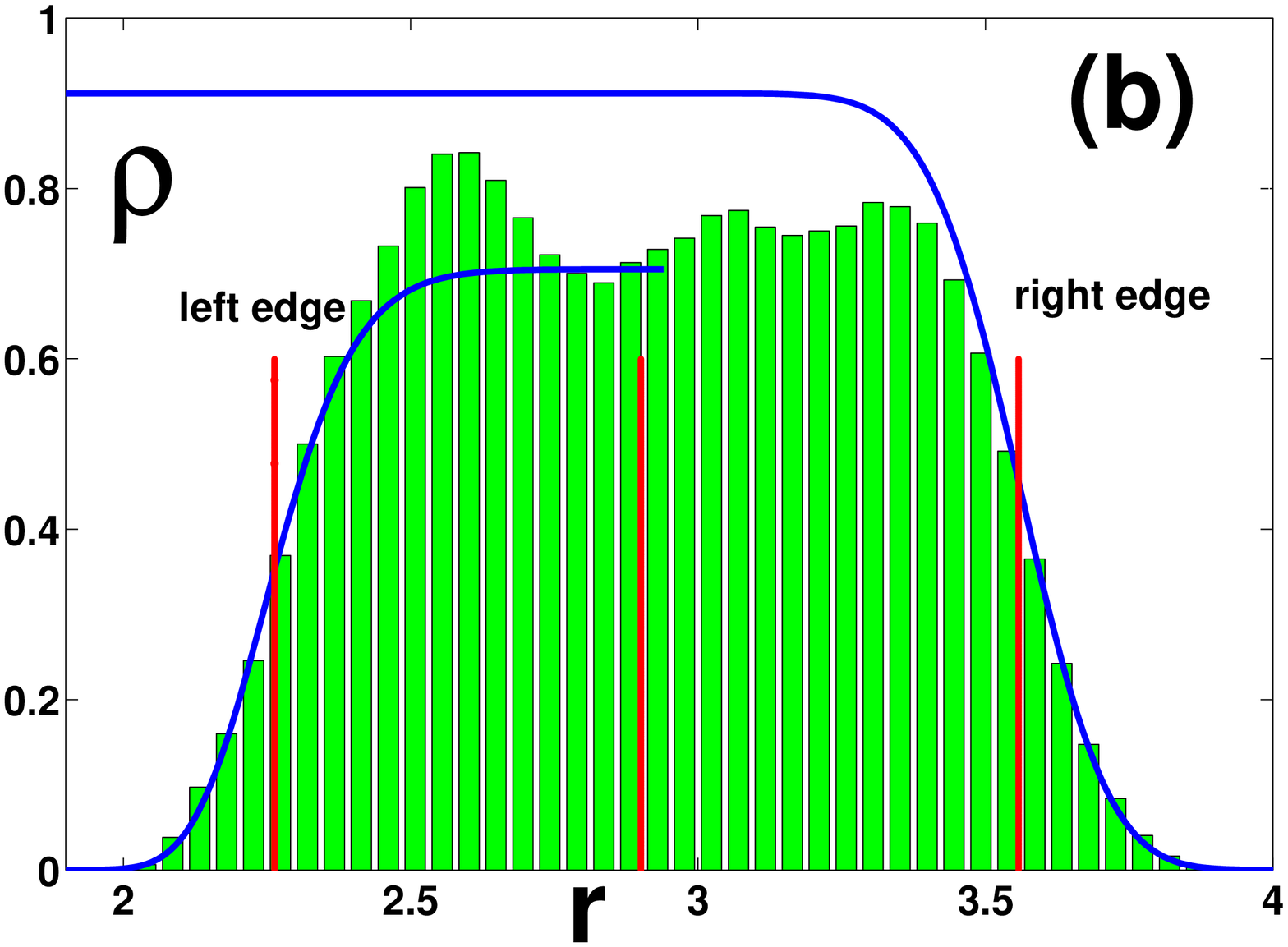} 
}\end{center}
\caption{\label{rescaled} \small The (non-rescaled) spectral densities of quantum maps $S^{(N)}\Lambda_{\bm \phi}$ v.s. asymptotics (\ref{Density}) (solid blue lines) for: (a) ``Doubly stochastic'' $10$-regular graphs with $N=1000$. Each vertex matrix $\sigma^v$ is fixed by  $10\times 10$ random unitary matrix.   (b)  ``damped'' De Brujin graph   with $N=2^7$ and $D_{i,j}=\delta_{i,j}f(2j\pi/N)$, $f(x)=3.2+\sin(x)+\sin(2x)+\sin(3x)$. The parameters are $\mu =0.1542$,  $\lambda_1=12.6578$ for the outer edge, and $\mu' =0.3133$, $\lambda'_1= 0.1952$ for the inner edge. The inner $\sqrt{1/\lambda'_1}$ and  outer radii  $\sqrt{\lambda_1}$ are depicted by vertical solid (red) left and right lines.  The (red) line in the middle shows the  mean value of  $\log f(x) $, where $\rho$ clusters at  $N\to\infty$. 
    }
\end{figure}

 {\it Comparison with numerics.}  Before  turning to the derivation of eqs.~(\ref{Density}, \ref{FormFactor}), let us consider several examples. {\sl (A) ``Connectivity'' graphs.} For a   graph $\Gamma$ take   $S$ be its connectivity matrix i.e., each  element of $S$ is either $0$ or $1$.  This choice is of a special interest due to the connection with the problem of length degeneracies in metric graphs~\cite{GO2013}. For the case of d-regular graphs it is straightforward to see that $\lambda=d$,  $\X=\bar{\X}= \mathds{1}$  implying   $\mu=d-1$.  The  comparison of (\ref{Density}, \ref{FormFactor}) with numerics for such a graph is shown on fig.~\ref{figwide} a,b.
{\sl (B) ``Doubly stochastic" graphs, fig.~\ref{rescaled}~a.} Let $S$ be such that $F$ is a doubly stochastic matrix i.e., $\sum_iF_{i,j}=\sum_jF_{i,j}=1$. This can be achieved, for instance, by taking  $\sigma^v_{i,j}=|u^v_{i,j}|$, where  $u^v$ are arbitrary unitary matrices.  It is known that these  matrices almost surely   satisfy the required spectral gap condition~\cite{Be2001}. As in the previous example  the highest eigenvector $\chi$ of $F$ is uniform, while $\lambda=1$. This gives  $\mu=\frac{1}{N}\tr(SS^{\dagger})^2-1$ (compare with the result of \cite{B2010} for RME with unitary invariant measures).
 {\sl (C) ``Damped'' quantum maps}  were suggested in~\cite{NS2008} as 
 toy models for open quantum  systems. They are represented as products $U_{\M}\cdot D$, where $N\times N$ unitary matrix  $U_{\M}$  is a quantization of a classical map $\M$ and ``smooth'' diagonal matrix $D$ introduces ``absorption''.   Here we checked  a particular case of Walsh quantized baker's map whose    quantization for $N=2^p$ can be written as
$U^{(p)} \Lambda_{\bm \phi}$, where $U^{(p)}_{i,j}=\frac{1}{\sqrt{2}}(\delta_{i,2i-1\mod N}-\delta_{i,2i}+\delta_{i,2i- N})$ and $\bm \phi$ is arbitrary, see \cite{Gu2010}. The matrix  $S^{(p)}=U^{(p)} D$ can  in turn be interpreted as the scattering matrix for the De Brujin graph with an absorption. We  compared spectral density  of  matrices $S^{(p)} \Lambda_{\bm \phi}$  with (\ref{Density}) and found  good agreement for both inner and outer edges, see fig.~\ref{rescaled}b. Note that in this case the edge distribution  does not ``converge'', since, as numerics shows, the parameter $\mu^{(N)}$ slowly grows with $N$.  This  observation agrees   with the phenomenon of eigenvalue clustering  near a ``typical'' value  found  in~\cite{NS2008}.

 {\it Derivation  of eqs.~(\ref{Density}, \ref{FormFactor}).} By the definition the form-factor $\mathcal{K}(n)$ can be represented as the double sum over $n$-periodic trajectories $\gamma$ of  the graph:
\begin{equation}\label{formFactorSum}
\frac{N}{n}\mathcal{K}(n)=\langle|\sum_{\gamma} A_\gamma e^{\i({\bm n}_\gamma,{\bm \phi})}|^2\rangle_{\bm\phi}=\sum_{\gamma,\gamma'} A_\gamma A_{\gamma'}^* \delta_{{\bm n}_\gamma,{\bm n}_{\gamma'}}
\end{equation}
with ${\bm n}_\gamma$ being  an integer-valued $N$ dimensional vector, whose elements $n_b$ indicate the number of times $\gamma$ visits  the bond $b$ ($\sum_b n_b=n$).
The amplitudes $A_\gamma$ are products of the matrix elements $S_{ij}$ taken along the path $\gamma$ and include the multiplicity factors which are $1$ for  prime periodic orbits. Following the standard semiclassical prescription~\cite{HaakeBook}  we analyze first the ``diagonal'', $\gamma=\gamma'$, contribution in~(\ref{formFactorSum}). Leaving out only prime periodic orbits and assuming  long trajectory limit ($n\sim\sqrt{N}\gg1$):
\[\sum_{i_1\dots i_n}  \abs{S_{i_1 i_2}}^2 \dots \abs{ S_{i_n i_1}}^2=\tr F^n= 1+\mathcal{O}(N^{-\frac{1}{2}+\kappa})\]
where we used the condition (i) on the spectral gap of $F$.

\begin{figure}[htb]
\begin{center}{
\includegraphics[height=1.6cm]{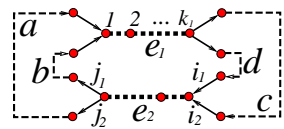} 
}\end{center}
\caption{\label{encounters} \small{
Diagram  of  periodic  orbits with two 2-encounters $e_1, e_2$ and four loops $a,b,c,d$  ($m=1$) contributing to (\ref{formFactorSum}). The two orbits of the same length are represented by the sequences $[ae_1ce_2be_1de_2]$, $[ae_1de_2be_1ce_2]$.}  }
\end{figure}

To calculate the next contribution one takes into account pairs of self-crossing trajectories. In each pair the partners $\gamma$, $\gamma'$ posses the same vector ${\bm n}_\gamma={\bm n}_{\gamma'}$, but traverse the bonds in different order, see fig.~\ref{encounters}. 
Note, that because  of broken time reversal symmetry  only trajectories with an even number of encounters make  contribution into (\ref{formFactorSum}). Furthermore, since $n$ is set to be of the same order as $\sqrt{N}$ only encounters with $2$ entering and $2$  exiting  loops should be considered. In a sharp contrast with  the unitary case (where the relevant scale is $n\sim N$), here the diagrams with $l$-encounters for $l>2$ contribute to the subleading order $o(N^0)$ only. 

The contribution from trajectories with $2m$ encounters can be split into a product of three factors  coming from:  encounters $\mathcal{N}_{\mathrm{enc}}$,    loops $\mathcal{N}_{\mathrm{loop}}$ connecting them  and  the combinatorics  $\mathcal{N}_{\mathrm{comb}}$. The latter takes into account all possible reconnections of loops and encounters i.e., the number of different diagrams. For the  diagrams with 2-encounters it is known~\cite{NBM2007} to be $\mathcal{N}_{\mathrm{comb}}=\frac{(4m)!}{2^{2m}(2m+1)!}$. 
Given $2m$ encounters there are $\frac{n^{4m}}{(4m)!}$ (to the leading order of $n$) choices to fix the lengths $\ell_1,\dots \ell_{4m}$ of the loops connecting them such that the total length is fixed $\sum_{i=1}^{4m}\ell_i=n-2\sum_{i=1}^{2m}(k_i-1)$, where $k_i$ is the length (i.e., the number of vertices) of the $i$-th encounter. The  contribution from all possible  paths  of the length $\ell\gg 1$ connecting   $j$th  and  $i$th bonds is given by $\sum_{i_1\dots i_{\ell-1}}  F_{i i_1}\dots F_{i_{\ell-1} j} = \bar{\chi}_i    \chi_j+\mathcal{O}(N^{-\frac{1}{2}+\kappa})$. This yields for the total  contribution from $4m$ loops with fixed entering and exiting bonds: 
$$
\mathcal{N}_{\mathrm{loop}}=\frac{n^{4m}}{(4m)!}\left(\prod_{r=1}^{2m} \bar{\chi}_{i_r} \chi_{j_r}\right) +\mathcal{O}(N^{-\frac{1}{2}+\kappa}).
$$

Given that  incoming  $(i_1,i_2)$ and  outcoming $(j_1,j_2)$ bonds of an encounter are fixed, the total contribution from all possible paths connecting them is 
\begin{eqnarray*}
\mathcal{N}_{\mathrm{enc}}^{(1)}&=&
(1-\delta_{i_1, i_2})(1-\delta_{j_1, j_2}) {S}_{i_1j_1} {S}_{i_1j_2}^* {S}_{i_2j_2} {S}_{i_2j_1}^*,\\
\mathcal{N}_{\mathrm{enc}}^{(k)}&=&\sum_{i,j}
(1-\delta_{i_1, i_2})(1-\delta_{j_1, j_2}) {F}_{i_1 i} {F}_{i_2 i} [Q^k]_{ij}{F}_{j j_1}{F}_{j j_2},
\end{eqnarray*}
 for encounters of the lengths  $k=1$  (containing a single vertex)  and   $k >1$, respectively.
Here $Q$ is an axillary matrix with the elements $Q_{ij}={F}_{ij}^2$. Combining these expressions with the factors  $\bar{\chi}_{i_1} \bar{\chi}_{i_2}\chi_{j_1} \chi_{j_2}$ from $\mathcal{N}_{\mathrm{loop}}$  and taking the sum  over the  indices gives for each encounter of the length $k$:
\begin{eqnarray*}
k=1:&&
\sum_{j_1,j_2,i_1,i_2} \bar{\chi}_{i_1} \bar{\chi}_{i_2}\chi_{j_1} \chi_{j_2}\mathcal{N}_{\mathrm{enc}}^{(1)} \\&&=
\tr\left[ (S\X S^\dag \bar{\X})^2-2(\bar{\X} \X)^2  + \bar{\X}^2 Q \X^2 \right];\\
k> 1: &&
\sum_{j,j_1,j_2,i,i_1,i_2} \bar{\chi}_{i_1} \bar{\chi}_{i_2}\chi_{j_1} \chi_{j_2}\mathcal{N}_{\mathrm{enc}}^{(k)}\\&&=
\tr\left[ \bar{\X}^2Q^k\X^2-2\bar{\X}^2Q^{k+1} \X^2  + \bar{\X}^2 Q^{k+2} \X^2 \right].
\end{eqnarray*}
 After summing up over all $k$, taking into account  $\mathcal{N}_{\mathrm{comb}}$  and the remaining combinatorial factor from $\mathcal{N}_{\mathrm{loop}}$ we arrive at
\begin{equation}\label{FormFactor1}
\mathcal{K}(n)=\frac{n}{N}\sum_{m=0}^\infty\frac{n^{4m}\mu^{2m}}{(2N)^{2m} (2m+1)!}+\mathcal{O}(N^{-\frac{1}{2}+\kappa}),
\end{equation}
which is the Tailor expansion of eq.~(\ref{FormFactor}). Finally, the spectral  density can be restored through the relationship
$$
\rho(r)=\frac{1}{\pi^2 r}\lim_{\varepsilon\to 0}\mathrm{Im}R_\varepsilon (r^{-1}),
 \quad R_\varepsilon (r)=  \sum_{n=1}^\infty
\left(re^{\i\varepsilon}\right)^n\mathcal{K}(n).
$$ 
by substituting (\ref{FormFactor}) and transforming the sum into integral. Applying the saddle point approximation to this integral in the regime $n\sim\sqrt{N}$ results in eq.~(\ref{Density}).

Formally  eqs.~(\ref{FormFactor},~\ref{Density})     can be also derived using the supersymmetry approach  of \cite{GA2004}. To this end the function $R_\varepsilon (r)$ is  represented  as the integral over supersymmetric ``fields". The result then follows by leaving out only  zero-mass mode. Contrary to the unitary case, however, even in the best case scenario of graphs with finite gaps the contribution of massive modes cannot be discounted on the basis of  a rough estimation suggested  in \cite{GA2004}.    
We defer the detailed  
discussion   of the supersymmetry approach to a later publication.

 In conclusion, we have shown that the spectral density and the form factor of the quantum map for strongly open quantum graphs  show the universal behavior at the edges of the spectrum at the scales of mean distance between eigenvalues. We conjecture that higher order spectral correlations exhibit  similar universality as well. In a sense our result can be seen as an extension of the well established universality  for closed quantum graphs. From the semiclassical point of view a strongly non-unitary case  differs in the  time scales involved: $\sqrt{N}$ rather than $N$. This results in  the exclusion of all diagrams with $l$-encounters for $l>2$.  In a transitional case with weak unitarity breaking, where $\mu^{(N)}\sim N^{-1}$,  these  diagrams must  be actually included since they contribute to the same order (in $n/N$) as (\ref{FormFactor1}). 

{\it Acknowledgments:}
 Financial support by SFB/TR12 and  DFG research grant Gu 1208/1-1  is gratefully acknowledged.

\end{document}